\documentstyle[epsfig]{elsart}
\begin{document}
\begin{frontmatter}
\title{\bf ``Super-radiance" and the width of exotic
baryons}
\author{N. Auerbach}
\address{School of Physics and Astronomy,Tel Aviv University, Tel
Aviv, 69978, Israel}
\author{V. Zelevinsky}
\address{NSCL, Michigan State University, East Lansing, MI 48824-1321, USA}
\author{A. Volya}
\address{Department of Physics,
Florida State University, Tallahassee, FL 32306-4350, USA}

\begin{abstract}

It is suggested that the narrow width of the recently observed
resonance $\Theta^{+}(1540)$ with strangeness $S=+1$ could be a
result of the super-radiance mechanism of the redistribution of
the widths of overlapping resonances due to their coupling through
common decay channels.

\end{abstract}
\end{frontmatter}

Fifty years ago, Dicke \cite{Dicke} introduced the notion of
``super-radiance" when discussing the formation of a short-lived
state in a radiant gas of $N$ identical two-level atoms. When the
gas is confined to a volume with a linear size smaller than the
wave length of the radiation, the atoms are coherently coupled
through the common radiation field. This leads to a formation of
the many-body state that has a width $\sim N\Gamma_{a}$, where
$\Gamma_{a}$ is the radiation width of an isolated individual
atom. This is the ``super-radiant" (SR) state observed in the
transmission of the laser pulse through the medium. The SR state
gets its width at the expense of other states of the system, which
are ``robbed" of their decay widths and become very narrow.

The phenomenon of the SR is in fact of a more general nature
\cite{sok,rot,ann,Izr,VZ}. The formalism can be derived with the
use of the Feshbach projection method \cite{Fesh,MW}, and in some
situations the SR state can be interpreted as a doorway for the
continuum coupling although we need to stress that the
conventional notion of the doorway state refers to the
collectivization of strengths in intrinsic dynamics that may or
may not coincide with the width collectivization and segregation
of direct and compound processes in the SR dynamics.  From the
late eighties, the ideas related to this concept have been applied
to many fields, including atomic physics \cite{auto}, molecular
physics \cite{p-v}, and condensed matter physics \cite{Pich}. In
low-energy nuclear physics
\cite{KR,alaga,savin,SB,Brent,Richt,BJW,VZ1} the main interest was
associated with resonances embedded in the continuum, collective
dynamics of giant resonances and description of loosely bound
nuclei. Applications to intermediate-energy nuclear physics
\cite{Auer} include meson resonances \cite{SB,Brent,Auer},
nucleon-antinucleon states \cite{Auer} and, more recently,
$\Delta$-isobar resonances in nuclei \cite{AZ}.

In many examples, including the last two cases, the emphasis was
placed not only on the wide SR state, but also on the fact that
the mechanism of the SR theory creates narrow resonances
superimposed on a background formed by the broad SR state leading
in this way to the separation of fast (direct) processes from the
long-lived structures. The consequences of this were examined and
it was suggested that this mechanism can explain the existence of
narrow autoionizing states in atoms and narrow resonances in a
number of strongly interacting hadronic systems, such as
dibaryons, hypernuclei etc.

\begin{figure}
\begin{center}
\epsfxsize=13.0cm \epsfbox{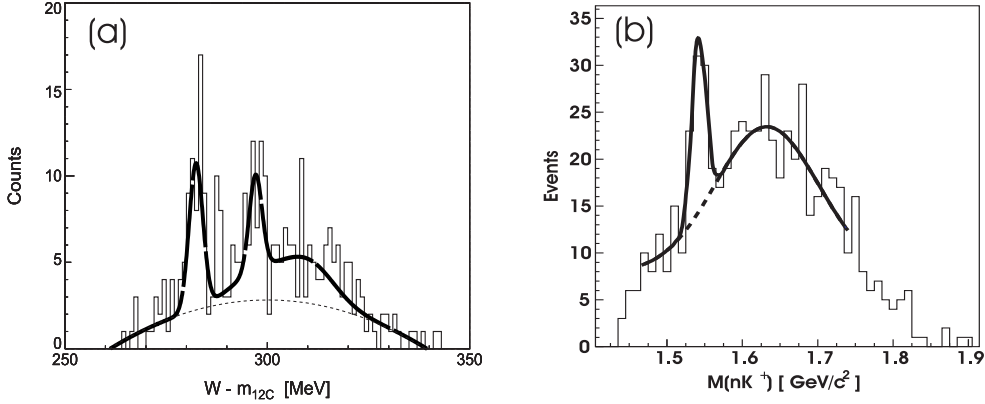}
\end{center}
\caption{\label{superradiance}Possible manifestation of the
superradiance in experiments. Figure in panel (a) is taken from
[20], it shows the observed counts in  $^{12}$C$(e,e'p\pi)^{11}$C
experiment as a function of energy. Panel (b) is taken from [26].
In both cases the narrow peaks are clearly seen on the background
of a broad SR state. The thick solid lines in both panels indicate
the experimental fit to the observed cross section, see the above
references for details.}
\end{figure}

The case of $\Delta$-resonances \cite{AZ} is especially relevant
for our discussion. In a $^{12}$C$(e,e'p\pi^{-})^{11}$C experiment
\cite{Mainz} at the Mainz microtron, at the energy near the
$\Delta$-isobar excitation narrow resonances (several MeV wide) in
the $\Delta-^{11}$C system were observed  superimposed on a broad
peak with the width $\sim 100\,$ MeV, see Fig.
\ref{superradiance}(a). In Ref. \cite{AZ} the broad peak was
interpreted as the SR state coherently formed out of the $\Delta
N^{-1}$ ($\Delta$ particle - nucleon hole in $^{11}$C)
configurations, while the narrow states corresponded to the
remaining, long-lived combinations of $\Delta N^{-1}$ excitations.

Before proceeding with another application, we present a brief and
simplified account of the SR state theory. Using a standard
projection technique, we consider a set of states $|q\rangle$
(termed ``internal") and a set of few ``external" decay channels
$|c\rangle$ \cite{Fesh,MW}. We refer to the spaces containing
these states as $\{q\}$ and $\{c\}$, respectively, whereas we
denote the projection operators onto these spaces as $Q$ and $P$.
The internal states $q$ with the same quantum numbers, such as
spin, isospin, parity, strangeness, may couple to each other via a
Hermitian interaction matrix $V_{qq'}$. Apart from that, in the
presence of decay channels, the intrinsic states can also interact
indirectly through the coupling $\langle q|V|c\rangle\langle
c|V|q'\rangle$ via the channel states $c$, the analog of the Dicke
interaction through the common radiation field. The effective
Hamiltonian ${\cal H}$ at given energy $E$, which belongs to the
continuum, can be put in the symbolic form with obvious notations,
\begin{equation}
{\cal H}_{qq'}=H_{qq'}+H_{qP}G^{(+)}(E)H_{Pq'},
                                              \label{1}
\end{equation}
where $H$ includes bare energies of the internal states and their
direct Hermitian interaction; the propagator through the
intermediate states $|c;E\rangle$,
\begin{equation}
G^{(+)}(E)=\frac{1}{E^{(+)}-H_{PP}},            \label{2}
\end{equation}
has to be taken at $E^{(+)}=E+i0$.

The effective intrinsic Hamiltonian (\ref{1}) is non-Hermitian
since the presence of decay thresholds $E^{c}_{t}$ creates an
imaginary part at energy above threshold. The real part of the
propagator in eq. (\ref{2}) (principal value) describes the
renormalization of the Hermitian interaction by the coupling
through all (open and closed) channels, while the imaginary part
of ${\cal H}$, $(-i/2)W_{qq'}$ with
\begin{equation}
W=2\pi\sum_{c({\rm open})}V|c\rangle\langle c|V,    \label{3}
\end{equation}
comes from the $\delta$-functions in the propagator corresponding
to the on-shell decay into the channels open at given energy.

The main features of the SR mechanism can be illustrated by the
simplest example of a single open channel $c_{0}$ and a diagonal
and degenerate real part of the effective Hamiltonian ${\cal H}$,
when the matrix elements of ${\cal H}$ have the form
\begin{equation}
{\cal H}_{qq'}=\epsilon\delta_{qq'}-\frac{i}{2}A_{q}A_{q'}^{\ast},
                                                   \label{4}
\end{equation}
where $A_{q}=\sqrt{2\pi}\langle q|V|c_{0}\rangle$ are the decay
amplitudes. The anti-Hermitian part here has a separable form,
similar to the classic model \cite{BB} for giant resonances, with
the only difference that the collectivization now occurs along the
imaginary axis via coupling through the continuum rather than
along real energy axis through multipole-multipole interaction
\cite{alaga}. With the rank of the matrix (\ref{4}) equal to 1,
all its eigenvalues are zero except for one (denoted by $r$) that
is equal to the imaginary part of the trace ${\cal H}$ and
accumulates the total summed width of all original states. For the
dimension $N$ of the internal space, the $N$ complex eigenvalues
of ${\cal H}$ are
\begin{equation}
{\cal E}_{n}=E_{n}-\frac{i}{2}\Gamma_{n}=\epsilon-i\pi\sum_{q}
|\langle q|V|c_{0}\rangle|^{2}\delta_{nr}\equiv
\epsilon-\frac{i}{2}\sum_{q}\Gamma_{q}\delta_{nr}. \label{5}
\end{equation}
We see in this schematic model that the coupling of the  states to
a single decay channel leads to the formation of a special state
that has a width equal to the sum of widths of all $N$ original
states, while the remaining $N-1$ states are stripped of their
decay widths and become stable. The broad state is just the SR
one. The separable form of the continuum interaction is not
arbitrary being dictated by the unitarity of the scattering
matrix.

In a more realistic situation, when the intrinsic energies
$\epsilon_{q}$ are not degenerate but their spacings are small
compared to widths, $\Delta\epsilon= |\epsilon_{q}-\epsilon_{q'}|<
\Gamma_{q}$ (overlapping resonances), a wide SR state still
appears but the rest of the states acquire their own widths, which
are much smaller than the width of the SR state (for more details
see Refs. \cite{sok,ann,Auer,AZ}). It was shown that the sum of
the widths, $\tilde{\Gamma}$, of all long-lived (``trapped")
states is given by
\begin{equation}
\tilde{\Gamma}\approx \frac{4(\Delta E)^{2}}{\Gamma},  \label{6}
\end{equation}
where $\Delta E$ is the energy spread of the $N$ internal states,
and $\Gamma=\sum_{q}\Gamma_{q}$.

As the number $k$ of open channels increases, so does the number
of wide states. However, the width segregation becomes less
pronounced because of more or less random phase interference of
the amplitudes for different channels in the imaginary part of the
Hamiltonian, eq. (\ref{3}), that reduces the effects of coupling
through the continuum.

The narrow resonances observed in the above mentioned
$^{12}$C$(e,e'p\pi)^{11}$C experiment \cite{Mainz} appear in the
SR theory in a natural way because the dynamics of the
$\Delta_{33}$-isobar in nuclei satisfies the validity conditions
of such theory. The wide peak around the $\Delta_{33}$ mass can be
interpreted \cite{AZ} as the SR state formed of a set of $\Delta
N^{-1}$ configurations, and the narrow states are the remaining
configurations of the same class with the same quantum numbers but
with decay probabilities strongly reduced by the SR mechanism.

Referring to the universality of the SR mechanism, we want to
point out in this paper that in the recent experiments
\cite{1,2,3,4,3a,5}, in which evidence was presented for a narrow
strange ($S=+1)$ baryon resonance around $E=1540\,$MeV, similar
conditions may occur as in the $^{12}$C$(e,e'p\pi)^{11}$C
experiment in the $\Delta_{33}$ region.

In various experiments one observed the width of this resonance,
coined as $\Theta^{+}$, or $Z^{+}$, smaller than 25 MeV; the
resonance is situated on top of a very broad (over 100 MeV)
background peak, Fig. \ref{superradiance}(b). In order to avoid a
contradiction with available data on kaon-nucleon scattering, the
actual width of the narrow resonance should be of the order of MeV
\cite{arndt,Haiden}. It is difficult to understand the existence
of this very narrow resonance structure in the framework of the
usual mechanism for decay of baryonic resonances, whether
non-exotic or exotic. The narrow width of the $S=+1$ exotic baryon
indicates that the decay into the $KN$ continuum is quenched
either by selection rules, for example if the $\Theta^{+}$ is an
isotensor \cite{isoten}, or by some special dynamical features.

Considering the Osaka experiment \cite{1} with $^{12}$C as a
target and the ITEP experiment \cite{2} with the Xe nucleus as a
target, the immediate thought would be that a mechanism similar to
that in the $^{12}$C$(e,e'p\pi)^{11}$C reaction is at work here
and the narrow width of the resonance is a result of a many-body
nuclear effect. The $\Delta N^{-1}$ configurations would be
replaced by the $\Theta^{+}N^{-1}$ ones and the $\pi N$ continuum
channel by the $K^{+}n$, or $K^{0}p$ in the ITEP experiment,
channel. The energy spacings between the various
$\Theta^{+}N^{-1}$ configurations with the same quantum numbers
coupled to the $K^{+}n$ or $K^{0}p$ channel are of the order of
$\hbar\omega_{N}$ or $\hbar\omega_{\Theta}$, which are the
spacings between the energies of major orbits of the nucleon or
$\Theta^{+}$-particle in the field of the nuclear core. These
spacings are typically of the order of 10-15 MeV (and can be
reduced by the larger mass of $\Theta^{+}$), i.e. much smaller
than the expected decay width into the kaon-nucleon continuum.

The CLAS experiment \cite{3,3a}, in which the deuteron is used as
a target, and the SAPHIR experiment \cite{4} with a hydrogen
target seemingly do not support the idea that the
$\Theta^{+}N^{-1}$ configurations are the cause for the quenched
decay of the resonance. However, as we explain, the SR state
mechanism may still be the cause for the quenching, and there is
not much difference in these cases compared to the experiments on
a complex nucleus as C or Xe.

The simplest scenario would be to consider the $K^{+}n$ system
($\bar{s}u+ddu$ quarks) to be nonrelativistic and to form a
quasimolecule. Molecular-like structures for non-strange
pentaquarks were discussed by Iachello \cite{iach}. The two
particles interact via an attractive potential of a typical range
of 1 fm producing a $p$-wave resonance at energy 100 MeV above
threshold (an $f$-wave resonance was considered in \cite{kahana}).
In Ref. \cite{Jaffe} the authors estimate that the width of the
resonance in such a potential is more than 175 MeV, the value
typical for strong decays of baryons in this mass region. A direct
calculation gives the $p$-resonance width 190 MeV for the well of
radius $a=1$ fm obtained at the needed energy, this required the
depth of the well to be $V=333$ MeV; the width of 327 MeV is
obtained for $a=2$ fm and $V=14$ MeV. These exact results can be
compared with the approximate formula for the $l$-wave resonance
width
\begin{equation}
\Gamma_l=\frac{2}{\mu a^2}\frac{(ka)^{2l+1}}{\left[(2l-1)!! \right
]^2} \left ( \frac{2l-1}{2l+1} \right ) \,,\quad l\ne 0.
                                                 \label{width}
\end{equation}
Although the fixed resonance energy $k^{2}/2\mu$ puts the relative
momentum of $K^{+}n$ system at $k=1.35$ fm$^{-1}$ that, with our
choices of the well size, does not satisfy the formal condition of
validity of Eq. (\ref{width}), $(ka)^2 <l(l+1)$, this formula
still turns out to give a good estimate and describe reasonably
well the scaling behavior of the $p$-wave width as a function of
$a$. An exact calculation for the $f$-wave resonance suppressed by
the centrifugal barrier gives the width of 4 MeV for $a=1$ fm that
required $V=1829$ MeV, and 49 MeV for $a=2$ fm with $V=316$ MeV.

However, the next excited (vibrational) states with the same
quantum numbers, that can be viewed as radial excitations of the
relative $K-N$ motion being analogous to the $\Theta^{+}N^{-1}$
excitations in heavier nuclei, are too high in energy and overlap
weaker although their widths are very large.

In addition to quasimolecular states, one should consider
many-quark bag dynamics in a system of five or even seven quarks
\cite{hepta} formed after the photon absorption by the original
proton. Among the intrinsic states there are groups with the same
quantum numbers, including ``normal" quark states, quark-gluon
states, paired states with singlet or triplet diquark(s), states
with pions and so on. If the decay width of each individual state
is much greater than the spacings, the unperturbed spectrum will
be that of overlapping resonances. Under such conditions, the
dynamics will be completely dominated by the coupling to the decay
channel, and the anti-Hermitian part of the Hamiltonian will be
the crucial factor. As a result, the SR state mechanism will give
rise to the observed spectrum with a very broad (several hundred
MeV) background peak, the SR state, and one or several very narrow
resonances with width of order of few MeV. For example, for two
overlapping resonances with $\Delta \epsilon \approx 20\,$MeV, and
the bare width of $\Gamma_{1,2}\approx 200\,$MeV, the estimate
above, eq. (\ref{6}), gives for the narrow resonance
$\Gamma_{nar}\approx 4\,$MeV. The molecular $KN$-state can play a
role of a doorway state \cite{ann,Auer,kahana} for the coupling to
the continuum and give the main contribution to the SR state after
the width redistribution. Then the direct (SR) processes are
spatially separated from the trapped states (compound processes),
similarly to what has been found in the open quantum wire model
\cite{ann}.

For a more quantitative discussion we consider a two-resonance
model discussed in detail in refs. (\cite{SB}) and (\cite{VZ1}) for
$\rho-\omega$ interference, halo nuclei and other applications. As
an example one can have in mind a quasimolecular state and a bag
state of a tightly bound five-quark system, although any two
intrinsic states with the same quantum numbers and a common decay
channel would be suitable as well. We assume that the unperturbed
energies of these states are $\epsilon_1$ and $\epsilon_2$,
respectively. The states are coupled to a single $KN$ channel (for
definitiveness we assume $p$-wave relative motion). The squared
amplitudes of these couplings $|A_1(E)|^{2}$ and $|A_2(E)|^{2}$
are greater than $\Delta \epsilon$ and have correct threshold
energy dependence. For the quasimolecular state the amplitude
$A_1(E)$ can be calculated assuming a square well potential; the
resulting behavior is well described by Eq. (\ref{width}), where
$\Gamma_1=|A_1(E)|^2$. The continuum coupling of a quark-bag state
in the near-threshold region is expected to have a similar
energy-dependence $A(E)\sim (E-E_{\rm t})^{3/4}$ being dominated
by the $p$-wave kinematics. Finally, both internal mixing and the
interaction through the continuum produce real coupling $v$
between these states. Due to a very different nature of the states
we expect this matrix element to be small. As a result, we come to
the effective non-Hermitian Hamiltonian,
\begin{equation}
{\cal H}=\left(\begin{array}{cc}
\epsilon_{1}-\frac{i}{2}\gamma_{1}&v-\frac{i}{2}A_{1}A_{2}\\
v-\frac{i}{2}A_{1}A_{2}&\epsilon_{2}-\frac{i}{2}\gamma_{2}
\end{array}\right),                        \label{heff}
\end{equation}
where $\gamma_{i}=|A_{i}(E)|^2$ ($i=1,2$).

The non-Hermitian, but symmetric eigenvalue problem (\ref{heff})
leads to the secular equations for the real and imaginary part of
resonance energy $E-(i/2)\Gamma$,
\begin{equation}
{E}^{2}-{E}(\epsilon_{1}+\epsilon_{2})-\frac{{\Gamma}}
{4}({\Gamma}-\gamma_{1}-
\gamma_{2})+\epsilon_{1}\epsilon_{2}-v^{2}=0,               \label{eq1}
\end{equation}
\begin{equation}
{\Gamma}=\frac{{E}(\gamma_{1}+\gamma_{2})-\gamma_{1}
\epsilon_{2}-\gamma_{2}\epsilon_{1}+2vA_{1}A_{2}}
{2{E}-\epsilon_{1}-\epsilon_{2}}.      \label{eq2}
\end{equation}
To avoid false solutions emerging because of the energy dependence
of the parameters, the two physical roots ${\cal E}_{\pm}=E_{\pm}
-(i/2)\Gamma_{\pm}$ are to be ``genetically" traced to the
original independently decaying states. This model provides a
transparent example of superradiance. As follows from Eqs.
(\ref{eq1}) and (\ref{eq2}) at energy above threshold, when both
decay amplitudes are non-vanishing and large, it is still possible
that an entire imaginary part is absorbed by one of the two
states. The condition for $\Gamma_{-}=0$ is
\begin{equation}
v(\gamma_1-\gamma_2)=A_1 A_2 (\epsilon_1-\epsilon_2).
                                               \label{sr}
\end{equation}
The parameters of pentaquark physics turn out to be close to
satisfying this condition.

\begin{figure}
\begin{center}
\epsfxsize=8.0cm \epsfbox{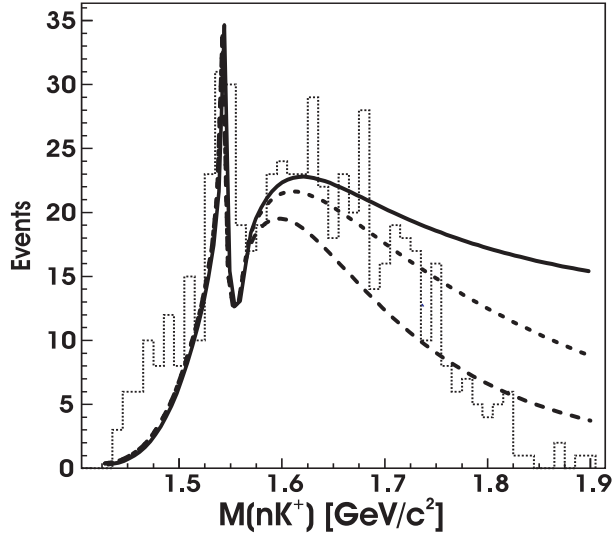}
\end{center}
\caption{\label{pq2l1} The experimental cross section, taken from
[26] is compared with a two-level model of superradiance. The
calculated cross section was approximately normalized to a number
of experimental events. The parameters of the model are discussed
in the text. Solid line corresponds to calculation where decay
amplitudes grow following Eq. (7), while for the dashed and the
dotted curves the high energy behavior is modified by a cutoff,
$\Lambda=300$  and 500 MeV, respectively. }
\end{figure}

For a qualitative comparison we consider what should be seen in
$KN$ scattering. In the two-resonance model with a single channel,
the resonance part of the scattering amplitude,
\begin{equation}
T(E)=\sum_{1,2} A_1^* \left (\frac{1}{E-{\cal H}}\right )_{12}
A_2, \label{scatt}
\end{equation}
where the denominator contains the full effective Hamiltonian
(\ref{1}), is given by
\begin{equation}
T(E)=\frac{E(\gamma_{1}+\gamma_{2})-\gamma_{1}\epsilon_{2}-
\gamma_{2}\epsilon_{1}-2vA_{1}A_{2}}{(E-{\cal E}_{+})(E-{\cal
E}_{-})}.                                    \label{32}
\end{equation}
Ignoring a non-resonant background, we obtain the cross section
$\sigma(E)=(\pi/k^2)|T(E)|^2.$ In Fig. \ref{pq2l1} the calculated
cross section is compared with the experimental graph. The
parameters selected for this example are $\epsilon_1=1535$ MeV,
$\epsilon_2=1560$ MeV, and $v=1$ MeV. For the solid curve, the
amplitudes of the continuum coupling for both states were assumed
to scale according to Eq. (\ref{width}), where the threshold
energy $E_{\rm t}=M_K+M_N=1432.3$ MeV.

The simple power-law scaling of decay amplitudes, Eq.
(\ref{width}), reflects the threshold behavior and breaks down as
the parameter $ka$ increases. Being not accounted for in the
calculations represented by the solid line, this leads to a large
cross section on a high energy side. To correct this we introduce
a cutoff energy $\Lambda$ that limits the validity of Eq.
(\ref{width}). For a quasimolecular state, the value of $\Lambda$
is about $200$ MeV; for a small quark bag, $\Lambda$ should be
higher. The dotted and dashed curves in Fig \ref{pq2l1} show the
results of the calculation with the same cutoff for both
amplitudes. The energy dependence was parameterized as $A(E)\sim
[E^3/(1+(E/\Lambda)^3)]^{1/4}$; at $E=1550$ MeV the bare widths of
the decoupled resonances are $\gamma_1=120$ MeV and $\gamma_2=60$
MeV. The precise form of the energy dependence is not important
(here for simplicity of notations we put the threshold at $E=0$).
The solution of the model defines the narrow resonance at required
energy 109 MeV above threshold with the width of 2.5 MeV and the
broad bump with the centroid at 120 MeV and width 178 MeV.

Certainly, the two-state model still should be considered merely
as an illustration. The parameters were not fit to the experiment,
being chosen just following the guiding physical principles so
that the set of our parameters is not unique. Additional factors,
such as possible isospin or/and flavor SU(3) violation, may also
play a role in the dynamics (according to the QCD sum rules
\cite{zhu}, the masses of the pentaquarks with strangeness +1 and
isospins 0,1 and 2 may be close that enhances their mixing). We
conclude that, due to the overlap of unstable intrinsic states,
the ``super-radiant" mechanism may produce the narrow peak(s) on
the broad background in exotic baryon systems with strangeness +1.
The dynamics leading to the redistribution of widths and formation
of the SR state along with long-lived trapped states are universal
in a sense that they are compatible with any intrinsic dynamics
being constrained by unitarity only. In various situations, as it
was discussed in \cite{alaga} and \cite{Auer,AZ}, the nature of
trapped intrinsic states can be very different. The observed
resonance widths can also be slightly different in different
nuclear environments.

The discussion with the theory group at the NSCL/MSU is highly
appreciated. V.Z. is grateful to F. Iachello for an interesting
discussion and the reference \cite{iach}; A.V. thanks S. Capstick
for a motivating discussion. The work was supported by the NSF
grant PHY-0244453 and in part by a grant from the US-Israel
Binational Science Foundation.

\end{document}